\documentclass[aps,pra,twocolumn,amssymb,floatfix]{revtex4}

\usepackage{graphicx}

\begin{document}

   \title{An Inverse--Problem Approach to Designing Photonic Crystals
          for Cavity QED}

   \author{JM Geremia}
   \email{jgeremia@Caltech.EDU}
   \homepage{http://minty.caltech.edu/MabuchiLab}
   \author{Jon Williams}
   \author{Hideo Mabuchi}
   \affiliation{Institute for Quantum Information, M.C. 12-33,
   California Institute of Technology, Pasadena, CA 91125}

   \date{\today}

\begin{abstract}

Photonic band gap (PBG) materials are attractive for cavity QED
experiments because they provide extremely small mode volumes and
are monolithic, integratable structures.  As such, PBG cavities
are a promising alternative to Fabry-Perot resonators. However,
the cavity requirements imposed by QED experiments, such as the
need for high Q (low cavity damping) and small mode volumes,
present significant design challenges for photonic band gap
materials. Here, we pose the PBG design problem as a mathematical
inversion and provide an analytical solution for a two-dimensional
crystal. We then address a planar (2D crystal with finite
thickness) structure using numerical techniques.

\end{abstract}

\maketitle

\section{Introduction} \label{Section::Introduction}

Engineering new materials to meet specific design objectives often
begins by trial and error.  New structures are repeatedly proposed
and characterized, and the results from each iteration are used to
further refine the design. This process continues until an
apparent optimum is achieved--- that is, when the incremental
modifications stop leading to improvements.

However, from an implementation standpoint, trial and error is
inefficient and costly, even when the process can be
computationally simulated. Conceptually, trial and error provides
little information about the quality of the optimum. This is
because the design space is often too large to permit an
exhaustive search. Therefore, it is common to fall back upon
physical intuition (that might not apply to the new material) to
guide the engineering process. Of course, this is not to say that
design-by-trial is ineffective, only that it lacks a certain
degree of rigor.

Going beyond incremental design procedures requires an
algorithmic, rather than intuitive, process.  In many cases,
posing the design problem as a mathematical
inversion\cite{Tihkonov,Geramb} can provide an assessment of the
resulting optimum.  Ideally, algorithmic searches might uncover
alternatives in the design space that physical intuition failed to
recognize.  However, such an unconstrained optimum structure might
prove too difficult to manufacture, in which case, the inversion
optimization can be restricted to account for limitations in the
fabrication capabilities.  Both alternatives are beneficial.  The
unconstrained inversion provides an indication of the absolute
optimal performance of the material, while the constrained
inversion produces the best structure that can actually be
constructed.

\vspace{2ex}

Engineering the optical properties of photonic band gap (PBG)
structures\cite{Joannopoulos1995, Yablonovitch1991, Scherer1999,
Plihal1991} is a process that can benefit from inversion
techniques.  Here, the objective is to tailor the electromagnetic
modes of the crystal by adjusting its spatially dependent
dielectric function.  Specifically, by introducing a defect into
an otherwise periodic crystal, it is possible to produce localized
electromagnetic fields\cite{Loncar2000}. Both the spatial and
temporal properties of these \textit{cavity modes} are affected by
the geometry of the defect.

The ability to localize light fields has made photonic crystals
attractive for experiments in cavity QED\cite{Vuckovic2001,
Vuckovic2000} and the quantum information
sciences\cite{Mabuchi2000,John1999}.  PBG cavities offer a number
of advantages that make them an attractive alternative to
Fabry-Perot resonators.  Most notably, a small cavity mode volume
is an important factor in achieving the strong coupling limit
between trapped atoms and the cavity light field.  Photonic
crystal cavities are capable of mode volumes on the order of the
cubic wavelength of the light\cite{Vuckovic2000, Yoshie2001}.

Modern lithographic techniques should enable the integration of
PBG structures with micron-scale magnetic traps for neutral
atoms\cite{Weinstein1995}. Atom trapping experiments, however,
pose non-trivial design challenges for photonic crystals.
Practical concerns in cavity QED experiments, such as atom
delivery and confinement, strongly suggest using planar photonic
crystals\cite{Miyai2001, Smith2000, Noda2000, Villeneuve1998,
Yoshie2001, Painter1999, Loncar2000, Johnson2001} (two-dimensional
lattices with finite thickness) rather than full 3D materials.
However, two-dimensional PBG crystals only provide incomplete,
\textit{quasi-3D} light trapping.  While well confined within the
lattice plane, the cavity field can decay in the out-of-plane
direction\cite{Vuckovic2001, Kawai2001,Painter1999a} by coupling
to radiated modes.  Radiation loss should generally be the most
significant decay mechanism in planar photonic
cavities\cite{Vuckovic2001}. Therefore, maximizing the cavity
quality factor, $Q=\Delta\omega_j / \omega_j$, requires that this
radiation loss be minimized. Simultaneously, cavity QED
experiments require that the cavity mode function have high
relative field strengths in vacuum regions (as opposed to inside
the semiconductor) that are accessible to trapped atoms.
Otherwise, the atomic system will not couple strongly to the
cavity field. Additionally, these criteria must be met without
sacrificing mode volume, \textit{i.e.}, by delocalizing the defect
field.

There has recently been considerable progress toward PBG cavities
that display the necessary properties for QED.  Numerical design
work performed in the Scherer group\cite{Vuckovic2001,
Vuckovic2001a} has identified planar photonic
crystals\cite{Sherer1998, Painter1999, Painter1999a} with mode
volumes on the order of the cubic wavelength of the light and
cavity Q factors $\sim$10$^4$, which is sufficient for strong
coupling. However, with these structures, it is difficult to know
if they are true optima. Furthermore, if they are not optimal, it
is unclear how to further improve them, \textit{i.e.}, whether
small modifications or large changes in the crystal would be
needed.

This paper poses the photonic crystal design process as an inverse
problem in order to provide an algorithmic optimization. That is,
we represent the defect dielectric function as \textit{resulting
from the design requirements}, rather than proposing a defect
geometry and then characterizing the crystal to observe its
properties.  To the best of our knowledge, this paper represents
the first attempt to treat photonic band gap materials in such a
manner.  We demonstrate that mathematical inversion leads to
photonic crystals structures not previously suggested by intuitive
or trial and error design techniques, and we also illustrate how
fabrication-imposed constraints can be placed on the inversion
optimization.

When solving inverse problems, there are two possible directions
to follow.  The first is to analytically treat a simplified model
that captures the relevant properties of the actual problem.
Analytic solutions are rarely possible for structures of arbitrary
(realistic) complexity; however, they provide a closed
mathematical description, and hopefully a better understanding, of
the design process.  In Section \ref{Section::Analytical} we
present analytic results for a pure (i.e. infinitely thick)
two-dimensional photonic crystal.  The second class of inversion
algorithms utilize numerical methods to treat more realistic
descriptions of the underlying physics. However, in exchange for
the more realistic model, it can be difficult to find global
extrema in the design space.  In Section \ref{Section::Numerical},
we employ numerical methods to treat a planar photonic crystal
membrane.

\section{Cavity Design as an Inverse Problem}

The relationship between the spatially dependent dielectric
function, $\kappa(\mathbf{r})$\footnote{Bold text in this paper is
used to represent vector quantities.}, and the properties of
interest, such as mode volume and Q, is a composition of two
individual components. First the electric and magnetic fields are
related to the reciprocal dielectric function, $\eta(\mathbf{r})=1
/ \kappa(\mathbf{r})$, through the Maxwell curl operator,
\begin{equation} \label{Equation::MaxwellCurl}
    \nabla \times \eta(\mathbf{r}) \nabla \times
    \mathbf{H}_j(\mathbf{r}) = \frac{\omega_j^2}{c^2}
    \mathbf{H}_j(\mathbf{r})
\end{equation}
where $\mathbf{H}_j(\mathbf{r})$ is the magnetic field for the
mode, $j$, with frequency, $\omega_j$.  It is often most
convenient to work with $\mathbf{H}$ because the resulting Maxwell
equation is Hermitian.  This provides no difficulty because the
electric field,
\begin{equation}
    \mathbf{E}_j(\mathbf{r}) = i \nabla \times
    \mathbf{H}_j(\mathbf{r}) / \omega_j \epsilon_0
    \kappa(\mathbf{r})
\end{equation}
can always be found from the magnetic field.

The second relationship then connects the photonic crystal's
electromagnetic modes to its characteristics. In some cases, the
property of interest can be directly expressed.  This is true for
the volume\cite{Berman},
\begin{equation} \label{Equation::ModeVolume}
    \mathcal{V}_j = \int | \psi_j(\mathbf{r}) |^2 d\mathbf{r}
\end{equation}
where $j$ is the mode index and $\psi(\mathbf{r})$ is defined,
\begin{equation} \label{Equation::VolumeFunction}
    \mathbf{H}_j(\mathbf{r}) = \mathbf{H}_{0,j} \psi_j(\mathbf{r})
\end{equation}
by choosing $\mathbf{H}_{0,j}$ such that $\psi_j$ is max-one
normalized, $|\psi(\mathbf{r})|_{max}=1$\footnote{In some cases,
the polarization may be position dependent, leading to
$\mathbf{H}^0_j=\mathbf{H}^0_j(r)$}. It is also possible to
directly express the magnitude of the field at the location,
$\mathbf{r}_a$, of the trapped atom,
\begin{equation}
    I_j = | \mathbf{H}_j(\mathbf{r}_a=0) |^2
\end{equation}
where $\mathbf{r}_a=0$ can always be achieved by a suitable choice
of axes.

For other characteristics, such as the cavity Q, it may be too
difficult, or inappropriate, to analytically express the property
as a function of the electromagnetic modes.  For the cavity Q, it
is more convenient to work with some other measure, $L$,
\begin{equation}
    Q_j \sim L_j[ \mathbf{H}_j(\mathbf{r}) ]
\end{equation}
that acts as a proxy for an actual calculation of the cavity Q.
This measure must display the property that maximizing it
simultaneously maximizes the Q (this will be discussed in greater
detail in Section \ref{Section::2DOptimization}).

\subsection{Inversion Cost Functional}

In all these cases, the fundamental relationships that connect the
properties of the electromagnetic modes to the reciprocal
dielectric function, $\eta(\mathbf{r})$, remain implicit,
\begin{equation}
    Q_j \rightarrow Q_j[\eta(\mathbf{r})], \quad \mathcal{V}_j \rightarrow
    \mathcal{V}_j[ \eta(\mathbf{r})], \quad I_j \rightarrow I_j[\eta(\mathbf{r}) ]
\end{equation}
Here, the notation, $[\cdots]$, represents the fact that the
quantities are complicated functionals of their input. This is
because Eq.\ (\ref{Equation::MaxwellCurl}) has been buried inside
them. Therefore, evaluating the functionals for any given crystal
entails solving Maxwell's equations and then computing the
property from the resulting modes.

Nonetheless, with these implicit functionals in hand, it is
possible to formally state the inverse problem by defining a cost
functional,
\begin{equation} \label{Equation::FormalCost}
    \mathcal{J}[ \eta(\mathbf{r}) ] = Q_m[ \eta(\mathbf{r}) ] +
    \beta_I I_m[\eta(\mathbf{r})] -
    \beta_{\mathcal{V}} \mathcal{V}_m[\eta(\mathbf{r})]
\end{equation}
evaluated for the appropriate cavity mode, $j=m$\footnote{In this
paper, the index $j$ refers to any mode supported by the photonic
crystal, while $m$ is the cavity mode}. $\beta_I$ and
$\beta_{\mathcal{V}}$ are scalars that balance the relative
importance of the various terms in the cost. Solving the inverse
problem is accomplished by optimizing $\mathcal{J}$,
\begin{equation} \label{Equation::FormalOptimization}
    \eta^* = \max_{\eta(\mathbf{r})}
     \mathcal{J}[\eta(\mathbf{r})]
\end{equation}
over the possible structures (indexed by their dielectric
function).

Of course, Eq.\ (\ref{Equation::FormalCost}) is deceptively
simple--- all the details of solving the inverse problem have been
relegated to the $[\cdots]$ notation.  However, there is an
important advantage to such an abstraction.  Equation
(\ref{Equation::FormalCost}) provides a language for describing
the photonic crystal structure \textit{in terms} of the design
objective.  It also provides a description of the inversion that
is independent of the particular method used to solve Maxwell's
equations.  $\mathcal{J}[\eta(\mathbf{r})]$ is easily generalized
to design objectives other than the cavity Q and mode volume, and
it applies to 2D, planar, and full 3D materials.

In the next two sections, Eqs. (\ref{Equation::FormalCost}) and
(\ref{Equation::FormalOptimization}) are solved for specific
examples. First, an analytic approach is used to treat an
infinitely thick two-dimensional crystal. In this case, handling
the $[\cdots]$ calls for the majority of the effort.  But once
this is accomplished, the optimization is relatively
straightforward. The second example incorporates numerical methods
to treat a planar 2D crystal. Here, there is no struggle with the
notation--- we just write a computer program to compute the mode
volume and cavity Q.  However, the optimization is complicated by
the possibility of local minima.  In both cases, it is necessary
to identify the best choice for the weighting parameters,
$\beta_I$ and $\beta_{\mathcal{V}}$, in the cost functional.

\section{Analytical Inversion} \label{Section::Analytical}

In this section, the photonic crystal inversion is analytically
performed for a two-dimensional structure.  The 2D problem is
motivated by the fact that planar and 2D structures share many
similarities. Treating the 2D crystal allows a detailed
mathematical inversion and can provide insight into how to also
optimize a planar crystal.

The general inversion strategy is to solve a variational problem
by expanding the cavity field in the bulk crystal electromagnetic
modes.  It is therefore necessary to select a bulk 2D lattice with
a band gap surrounding the desired cavity resonance frequency
(such as an hexagonal array of holes with a suitable lattice
constant) prior to the inversion.  Once the electromagnetic modes
of the bulk structure are determined, it is possible to optimize
the cavity Q, field intensity at $\mathbf{r}_a$, and mode volume
over the bulk mode expansion coefficients.  This optimization
stage does not directly involve the defect dielectric function---
it identifies the optimal cavity field that can be produced using
the bulk crystal modes as a basis. Once these optimal expansion
coefficients are identified, the defect that produces the optimal
field is extracted by inverting the Maxwell curl equation, Eq.
(\ref{Equation::MaxwellCurl}).

\subsection{Bulk Crystal Modes}

Methods for solving the Maxwell equations for a two-dimensional
photonic crystal are well established\cite{Plihal1991,
Plihal1991a}. However, it is useful to briefly review the plane
wave expansion method in order to provide sufficient context for
solving the inverse problem.

The two-dimensional photonic crystal consists of a bulk medium
with index of refraction, $n_b$.  It is laced with a lattice of
infinitely deep cylindrical holes with radius, $r_h$, and index of
refraction, $n_h$.   This lattice is represented by the real space
vectors, $\{\mathbf{R}_n\}$, which point from the origin to the
centers of the cylinders.  All of the real space lattice vectors
lie in a plane.

Since the inverse dielectric function, $\eta_0(\mathbf{r})$, for
the bulk crystal is periodic, it is convenient to work with its
Fourier transform,
\begin{equation} \label{Equation::BulkFourier}
    \eta_0(\mathbf{r}) = \sum_G \eta_{\mathbf{G}} e^{i \mathbf{G} \cdot \mathbf{r}}
\end{equation}
where the reciprocal lattice vectors, $\mathbf{G}$, satisfy
$\mathbf{G} \cdot \mathbf{R}_n=2l\pi$, $l=1,2,\ldots$. Physically,
each reciprocal lattice vector is the wave vector of a plane wave
that shares the periodicity of the real space lattice.

As with the dielectric function, the bulk crystal electromagnetic
modes are periodic in the lattice.  In accordance with Bloch's
theorem\cite{AshcroftMermin}, the bulk crystal electromagnetic
modes can also be expanded in the reciprocal lattice vectors,
\begin{equation} \label{Equation::BulkModeBloch}
    \mathbf{H}_{n,\mathbf{q}}(\mathbf{r}) = \sum_{\lambda}
    \hat{\mathbf{e}}_{\lambda}
    \sum_G h_{n,\mathbf{q} + \mathbf{G}} e^{i (\mathbf{q} + \mathbf{G} )
    \cdot \mathbf{r} }
\end{equation}
where the mode is labelled by its wavevector,
$\mathbf{q}$,\footnote{In this paper, $\mathbf{G}$ always
represents a reciprocal lattice vector, $\mathbf{q}$ refers to a
wave vector confined to the first Brillouin zone, and $\mathbf{k}$
refers to any wave vector that is not necessarily on the
reciprocal lattice or in the Brillouin zone.} and band index, $n$.
The $\mathbf{\hat{e}}_{\lambda}$ are orthogonal polarization
vectors and the $h_{n,\mathbf{q}+\mathbf{G}}$ are the plane wave
expansion coefficients that produce the mode.

Calculating the bulk crystal modes is accomplished by solving the
Maxwell equation, Eq.\ (\ref{Equation::MaxwellCurl}), using the
form in Eq.\ (\ref{Equation::BulkModeBloch}).  This leads to wave
equations for the two possible polarizations,
\begin{equation}
    \sum_{\mathbf{G}^\prime} \eta_{\mathbf{G}-\mathbf{G^\prime}}
    (\mathbf{q}+\mathbf{G})\cdot(\mathbf{q} +\mathbf{G}^\prime)
    h_{n,\mathbf{q}+\mathbf{G}^\prime}=\frac{\omega^2_{n,\mathbf{q}}}{c^2}
    h_{n,\mathbf{q+\mathbf{G}}}
\end{equation}
for TE modes and,
\begin{equation}
    \sum_{\mathbf{G}^\prime} \eta_{\mathbf{G}-\mathbf{G^\prime}}
    | \mathbf{q}+\mathbf{G} | |\mathbf{q} +\mathbf{G}^\prime|
    h_{n,\mathbf{q}+\mathbf{G}^\prime}=\frac{\omega^2_{n,\mathbf{q}}}{c^2}
    h_{n,\mathbf{q+\mathbf{G}}}
\end{equation}
for TM modes.  Since the polarizations uncouple for a pure
two-dimensional crystals, it is possible to work with them
independently.  For the remainder of this section, we utilize TE
modes; however, the same inversion technique applies equally well
to TM modes.

\subsection{Defect Crystal Modes}

The cavity mode, $m$, can be expanded in the bulk modes,
$\mathbf{H}_{n,\mathbf{q}}(\mathbf{r})$, using wavevectors that
are confined to the first Brillouin zone (see for
example\cite{AshcroftMermin}),
\begin{equation} \label{Equation::BulkExpansion}
    \mathbf{H}_m(\mathbf{r}) =
    \sum_n \sum_{\mathbf{q} \in BZ} a_{n,\mathbf{q}}^{(m)}
    \mathbf{H}_{n,\mathbf{q}}(\mathbf{r})
\end{equation}
of the two-dimensional lattice.  This superposition is the reason
for working with the magnetic, rather than electric, fields. Since
Eq.\ (\ref{Equation::MaxwellCurl}) is Hermitian, the bulk modes
are complete.

The photonic crystal cavity can be described by introducing an
additional defect term, $\delta\eta(\mathbf{r})$, into the
reciprocal dielectric function,
\begin{equation} \label{Equation::DefectDielectric}
    \eta(\mathbf{r}) = \eta_0(\mathbf{r}) + \delta\eta(\mathbf{r})
\end{equation}
whose Fourier transform is given by,
\begin{equation} \label{Equation::DefectFourier}
    \delta\eta(\mathbf{r}) =  \int_{\mathbf{k}} d \mathbf{k}
    \delta\eta(\mathbf{k}) e^{i \mathbf{k} \cdot {\mathbf{r} } }
    \approx \sum_{\mathbf{k}}
    \delta\eta_{\mathbf{k}} e^{i \mathbf{k} \cdot {\mathbf{r} } }
\end{equation}
However, unlike the bulk lattice, the cavity is not periodic, so
the Fourier expansion must run over all wave vectors,
$\mathbf{k}$.  In practice, the integral is generally approximated
by a discrete sum that is then truncated to allow computation.

The coefficients, $a^{(m)}_{n,\mathbf{k}}$, are calculated by
substituting Eqs.\ (\ref{Equation::BulkExpansion}) and
(\ref{Equation::DefectDielectric}) into the Maxwell equation, Eq.\
(\ref{Equation::MaxwellCurl}).  This produces the matrix
eigenvalue equation,
\begin{equation}
    \sum_{n^\prime,\mathbf{q}^\prime}
    D^{(m)}_{n,\mathbf{q};n^\prime,\mathbf{q}^\prime}
    a^{(m)}_{n^\prime,\mathbf{q}^\prime} =
    \frac{\omega_m^2}{c^2}
    a^{(m)}_{n^\prime,\mathbf{q}^\prime}
\end{equation}
where
\begin{eqnarray}
   D^{(m)}_{n,\mathbf{q};n^\prime,\mathbf{q}^\prime} =
   \sum_{\mathbf{G,G^\prime}}
   h^*_{n,\mathbf{q+G}} \delta\eta_{\mathbf{q+G-q^\prime-G^\prime}}
   h_{n^\prime,\mathbf{q^\prime+G^\prime}} & & \\
   (\mathbf{q+G}) \cdot (\mathbf{q^\prime+G^\prime}) +
   \delta_{n,n^\prime} \delta_{\mathbf{q,q^\prime}}
   \frac{\omega_{n,\mathbf{q}}}{c^2} & & \nonumber
\end{eqnarray}
Here, it can be seen that the point defect couples all of the bulk
crystal modes.  However, the Fourier coefficients,
$\delta\eta_{\mathbf{k}}$, fall off quickly as the magnitude of
the wavevector increases.

\subsection{Cavity Mode Optimization} \label{Section::2DOptimization}

To optimize the inversion cost functional, $\mathcal{J}$, it is
necessary to express Eq.\ (\ref{Equation::FormalCost}) in terms of
the defect crystal modes by utilizing the expansion, Eq.\
(\ref{Equation::BulkExpansion}).  When performing an actual
inversion calculation, this is the point when it would be
necessary to select a bulk photonic crystal geometry, such as a
hexagonal lattice.  Once this has been done, the plane wave
representations of the optimization basis functions,
$\mathbf{H}_{n,\mathbf{q}} (\mathbf{r})$, can be computed.

Another important point is that the optimization is performed over
$\delta\eta(\mathbf{r})$, not the bulk lattice function,
$\eta_0(\mathbf{r})$.  In principle, this does not restrict the
optimization in any way.  The distinction between
$\eta_0(\mathbf{r})$ and $\delta\eta(\mathbf{r})$ is not
perturbative, so there is no requirement on the relative
magnitudes of the two functions.  However, in practice, it can be
practical to restrict the structure of the defect (for example, to
enforce radial symmetry) in order to limit the number of plane
waves (or equivalently, the number of reciprocal lattice vectors)
needed for Eq.\ (\ref{Equation::BulkModeBloch}) to converge.

\subsubsection{Cavity Mode Volume and Intensity}

Expressing the cavity mode volume in terms of the basis function
coefficients is straightforward,
\begin{equation} \label{Equation::ModeVolumeBasis}
    \mathcal{V}_m = \sum_{n^\prime,\mathbf{q}^\prime}
    \sum_{n,\mathbf{q}} a^{(m)*}_{n^\prime,\mathbf{q}^\prime}
    a^{(m)}_{n,\mathbf{q}} \langle
    \psi_{n^\prime,\mathbf{q}^\prime}(\mathbf{r}) |
    \psi_{n,\mathbf{q}}(\mathbf{r}) \rangle
\end{equation}
as is taking derivatives with respect to the coefficients,
\begin{equation}
    \frac{d\mathcal{V}_m}{d a_j} = \sum_{n,\mathbf{q}}
    a^{(m)}_{n,\mathbf{q}} \langle \psi_{j}(\mathbf{r}) |
    \psi_{n,\mathbf{q}}(\mathbf{r}) \rangle
\end{equation}
Here, the spatial functions, $\psi_j(\mathbf{r})$, are the
quantities defined in Eq.\ (\ref{Equation::VolumeFunction}) and
the inner product, $\langle \psi_{n^\prime,\mathbf{q}^\prime} |
\psi_{n,\mathbf{q}}\rangle$, denotes integration over the
real-space domain in which the mode volume is to be minimized.
This domain is arbitrary, provided that it is at least as large as
the photonic crystal to be used in the cavity QED experiment.

Similarly, the field intensity at the location of the trapped atom
is given by,
\begin{equation}
    I_m = \sum_{n^\prime,\mathbf{q}^\prime}
    \sum_{n,\mathbf{q}} a^{(m)*}_{n^\prime,\mathbf{q}^\prime}
    a^{(m)}_{n,\mathbf{q}}
    \mathbf{H}^*_{n^\prime,\mathbf{q}^\prime}(0)
    \mathbf{H}_{n,\mathbf{q}}(0)
\end{equation}
and the necessary derivatives with respect to the expansion
coefficients are also straightforward to find.  When performing an
inversion calculation, both the mode volume and intensity
functions can be further expanded in terms of the reciprocal
lattice vector plane waves.  Doing so leads to algebraic
expressions in the coefficients, $h_{n,\mathbf{q+G}}$, from Eq.\
(\ref{Equation::BulkModeBloch}).

\subsubsection{Cavity Q Factor}

It is not as clear how to represent the cavity Q in terms of the
basis functions.  The two-dimensional lattice is infinitely deep
and therefore does not permit any radiated modes.  However, this
does not prevent us from minimizing features of the
two-dimensional lattice that would promote out-of-plane loss were
the structure a planar crystal.  In other words, we wish to find
the two dimensional structure of a planar photonic crystal that
minimizes the out-of-plane loss by considering features computed
for a pure 2D structure.   Since the cavity Q cannot be directly
computed, it is necessary to define an auxiliary measure of field
decay that applies to the two-dimensional lattice.

The essential requirement of the auxiliary measure is that
optimizing it simultaneously maximizes the cavity Q for a planar
structure (where radiation loss can occur). It is possible to
identify such a measure by considering the physical nature of
radiative field decay in a planar photonic crystal. Out-of-plane
loss is the result of guided crystal modes coupling to
free-space\cite{Vuckovic2001} and frequency-wavevector pairs in
free space must lie within the light cone.  Therefore, bulk
crystal modes with frequency-wavevector pairs that lie below the
light line should not couple to free space because they undergo
total internal reflection\cite{Vuckovic2001a}.

Minimizing the contributions from bulk modes,
$\mathbf{H}_{n,\mathbf{q}}(\mathbf{r})$ that lie above the light
line reduces radiative cavity decay.  We chose to adopt the
following auxiliary function,
\begin{equation} \label{Equation::ComponentBasis}
      L = \sum_{n^\prime,\mathbf{q}^\prime}
    \sum_{n,\mathbf{q}} a^{(m)*}_{n^\prime,\mathbf{q}^\prime}
    a^{(m)}_{n,\mathbf{q}}
    \frac{\omega_{n^\prime,\mathbf{q}^\prime}\omega_{n,\mathbf{q}}}
    {qq^\prime}
\end{equation}
where $q_j$ is the magnitude of its corresponding wavevector,
$q_j=\|\mathbf{q}_j\|$.

\subsubsection{Analytic Optimization}

The photonic cavity design characteristics, Eqs.
(\ref{Equation::ModeVolumeBasis}-\ref{Equation::ComponentBasis}),
can be substituted for their respective terms in the cost
functional, $\mathcal{J}[\delta\eta]$.  Setting the derivatives of
the design properties with respect to the expansion coefficients
equal to zero produces a linear variational problem. In order to
insure that the mode functions remain properly normalized, it is
convenient to impose the constraint, $1 - \sum
|a_{n,\mathbf{q}}|^2 = 0$, as a Lagrange multiplier.  Maximizing
the resulting Lagrangian leads to a matrix eigenvalue problem,
\begin{eqnarray}
    & \sum_{n^\prime,\mathbf{q}^\prime} \left[
    \frac{\omega_{n^\prime,\mathbf{q}^\prime}\omega_{n,\mathbf{q}}}
    {qq^\prime}
    + \beta_I H^*_{n^\prime,\mathbf{q}^\prime}(0)
    H_{n,\mathbf{q}}(0) \right. & \\
    \label{Equation::AnalyticOptimum}
    & \left. - \beta_{\mathcal{V}} \langle
    \psi_{n^\prime, \mathbf{q}^\prime} |
    \psi_{n,\mathbf{q}} \rangle \right]
    a^{(m)}_{n^\prime,\mathbf{q}^\prime} = \Lambda
    a_{n,\mathbf{q}} & \nonumber
\end{eqnarray}
whose eigenvectors correspond to values of the expansion
coefficients, $a^{(m)}_{n,\mathbf{q}}$, that satisfy Eq.\
(\ref{Equation::FormalOptimization}).  The eigenvector
corresponding to the smallest eigenvalue is the best optimum.

In practice, it is necessary to select values for the weighting
parameters, $\beta_I$ and $\beta_{\mathcal{V}}$.  Although
appropriate values can be found by balancing the relative
magnitudes of the three terms in the cost, it is better to nest
Eq.\ (\ref{Equation::AnalyticOptimum}) within a second, numerical
maximization, over the $\beta_i$.   Although, performing the
$\beta_i$ optimization calls for repeating the eigenvalue problem,
possibly many times,  saving the values of the variational matrix
elements and simply rescaling them according to the particular
choice of the $\beta_i$ offers considerable computational savings.

\subsection{Extracting the Defect Dielectric}
    \label{Section::2DInversion}

With the optimal mode coefficients, $a^{(m)}_{n,\mathbf{q}}$,
known, the final component of the inversion is to extract the
defect dielectric, $\delta\eta(\mathbf{r})$, from the expansion
coefficients. Doing so involves inverting the Maxwell equations,
and can be accomplished by substituting Eqs.
(\ref{Equation::BulkFourier}), (\ref{Equation::BulkModeBloch}) and
(\ref{Equation::DefectFourier}) into Eq.
(\ref{Equation::MaxwellCurl}) and solving for the
$\delta\eta_{\mathbf{k}}$.

In simplifying the resulting expressions, it is necessary to make
use of the orthogonality of the bulk crystal mode functions.  It
is also helpful to let the mode wavevectors, $\mathbf{q}$, run
over multiple Brillouin zones.  Doing so leads to more manageable
equations because the summations are no longer restricted.  In the
end, the proper expressions can be obtained by folding the
equations back into the first Brillouin zone.  The details of the
derivation are provided in Appendix \ref{Appendix::Details}, and
the result is a linear system of equations,
\begin{equation} \label{Equation::2DSystem}
    \sum_{\mathbf{k}} D^{(m)}_{n,\mathbf{q};\mathbf{k}}
    \delta\eta_{\mathbf{k}} = a^{(m)}_{n,\mathbf{q}}
    \frac{\omega_m^2 - \omega_{n,\mathbf{q}}^2}{c^2}
\end{equation}
where the inversion matrix, $\mathbf{D}^{(m)}$, is given by
\begin{eqnarray}
   & D^{(m)}_{n,\mathbf{q};\mathbf{k}} =
   \sum\limits_{n^\prime} \sum\limits_{\mathbf{G},
   \mathbf{q}^\prime}
   a^*_{n^\prime,\mathbf{q}^\prime}
   h^*_{n,\mathbf{q+G}} h_{n^\prime,\mathbf{q^\prime+G-k^\prime}}
   & \label{Equation::2DInverseMatrix} \\
   & (\mathbf{q+G})\cdot(\mathbf{q+G-k^\prime}) & \nonumber
\end{eqnarray}
The matrix is indexed by the bulk modes, labelled by
($n,\mathbf{q}$), and the Fourier coefficients of the defect,
$\mathbf{k}$.

An important point to make is that the cavity resonance frequency,
$\omega_m$, enters into Eq. (\ref{Equation::2DSystem}) as a
parameter.  Solving the inversion requires specifying the cavity
frequency, which can take on any value within the bulk crystal
band gap. It should be expected that the best in-plane confinement
results from a cavity frequency, $\omega_m$, deep within the band
gap.  However, the resulting inverted defect dielectric function
is different depending on the choice of the resonance frequency.
Moreover, different choices of $\omega_m$ might lead to defects
that are easier to fabricate than others.  This property of the
inversion is one that likely requires further investigation.

\subsubsection{Computational Complexity}
\label{Section::Complexity}

Interpreting the computational complexity of the matrix equations
in Eq.\ (\ref{Equation::2DInverseMatrix}) is aided by considering
the $\mathbf{k}$ vectors as a sum of reciprocal lattice vectors
and Brillouin zone vectors, $\mathbf{k}=\mathbf{q+G}$.  This shows
that the defect is constructed using bulk crystal modes for
\textit{all} wavevectors that lie within the Brillouin zone.
Counting vectors in this manner allows the dimensions of the
inversion matrices to be determined.  Since Eq.\
(\ref{Equation::MaxwellCurl}) is Hermitian, the number of bands is
equal to the number of reciprocal lattice vectors, $N_G=\infty$.
Similarly, the number of bulk crystal modes needed to construct a
non-periodic defect is $N_q=\infty$. Therefore, the dimension of
the square matrix, $\mathbf{D}^{(m)}$, is a strictly countable
double infinity.

In practice, summations over reciprocal lattice vectors are
truncated to a finite number, $N_G$, which likewise limits the
number of bands for each wavevector to $N_G$.   The truncated bulk
mode basis is no longer rigorously complete. However, in practice,
the Fourier coefficients in Eq.\ (\ref{Equation::BulkFourier})
fall off quickly as $\| \mathbf{G} \|$ increases.   Therefore,
truncating the basis can be achieved without sacrificing accuracy
provided that Fourier expansions remain sufficiently accurate.
Retaining the proper dimension of $\mathbf{D}^{(m)}$, requires
that the number of defect Fourier wavevectors, $N_k$, also be
properly chosen.  The requirement that $\mathbf{D}^{(m)}$ remain
square requires that the number of distinct bulk modes in the
Brillouin zone, $\mathbf{q}$ (recall that
$\mathbf{k}=\mathbf{q+G}$) be exactly $N_G$.

The number of matrix elements in $\mathbf{D}^{(m)}$ scales as
$O(N_G^2)$.  However, the computational complexity of the sums
that must be evaluated when computing the matrix elements scale as
$O(N_G^3)$. Therefore, the overall time complexity for solving the
inverse problem is $O(N_G^5)$, and can prove to be computationally
demanding.  In practice, it is beneficial to employ approximate
methods for solving the linear system of equations\cite{Golub} if
the calculations are to be performed on a personal computer.

\subsection{Illustration: Symmetric Defect}
      \label{Section::SymmetricDefect}

As a first demonstration, we considered an hexagonal 2D photonic
crystal with a radially symmetric defect.  The bulk material index
of refraction was chosen to be, $n_b=3.4$, and the hole radius
was, $r_h/a=0.3$ (where $a$ is the lattice constant). The
rationale behind these parameters was based on competing factors.
It has been shown that large lattice holes lead to out-of-plane
loss due to scattering from the
edges\cite{Vuckovic2001,Vuckovic2001a}. However, the frequency
band gap decreases as the hole size is reduced, so $r_h$ must not
be made too small.  30\% of the lattice constant provides a good
compromise.
\begin{figure}[hb]
\caption{The
photonic crystal (A) and the cavity electric field (B) that result
from optimizing the Q, mode volume, and peak cavity intensity by
solving a two-dimensional inverse problem for a radially symmetric
defect. The resulting cavity hole was reduced in radius to
$r_c=2.1/a$, and the neighboring holes were also reduced in size
and displaced outward.  The most significant feature of the
optimized structure is that the index of refraction of the bulk
crystal holes was increased to $n_h=1.9$.
\label{Figure::SymmetricMode}}
\end{figure}

\begin{figure}[ht]
\caption{
   Extracted reciprocal dielectric function, $\eta(\mathbf{r})=
   1/\kappa(\mathbf{r})$,
   found by solving the inversion problem for a two-dimensional
   hexagonal lattice with a radially symmetric defect.  The most
   distinctive feature of the mathematically optimal photonic
   crystal is that the bulk lattice air holes have an index
   of refraction $n_h>1$.
\label{Figure::SymmetricEta}}
\end{figure}

The dielectric index of refraction, $n_b=3.4$, was chosen because
it is typical of the semiconductor materials that might be used to
incorporate photonic crystals into atom trapping experiments
(\textit{e.g.}, Al$_.3$Ga$_.7$As). The essential property of the
bulk dielectric material is that it not absorb light around the
atomic transition frequency, for example, 852 nm for Cs.

For the purpose of the calculations, the photonic crystal lattice
was truncated to five layers surrounding the center defect at
$\mathbf{r}=0$.  We found that this number of layers provided a
sufficient description of the properties of the photonic crystal
without exceeding the computational power of a typical desktop
computer. The coefficient optimization was performed by adjusting
the parameters, $\beta_i$, in $\mathcal{J}$ until the best maximum
was achieved.  Parameter optimization was performed using a
conjugate-gradient search algorithm\cite{NumericalMethods} over
the $\beta_i$.  The search required solving the eigenvalue problem
in Eq. (\ref{Equation::AnalyticOptimum}) for different scaling
parameters 28 times.

Solving the linear system of inversion equations in
\ref{Section::2DInversion} produced the crystal and associated
cavity mode shown in Figure \ref{Figure::SymmetricMode}.  Here,
the 2D mode volume was $\sim$$\lambda^2/4$.  The location and size
of the photonic crystal holes was determined from a contour plot
of the reciprocal dielectric function, Figure
\ref{Figure::SymmetricEta}.  The contour half-way between the
minimum and maximum values of $\eta(\mathbf{r})$ was adopted in
order to eliminate the small peak oscillations.  These
oscillations were the result of truncating the Fourier expansions
for $\eta_0(\mathbf{r}$ and $\delta\eta(\mathbf{r})$.

The photonic crystal shown in Figure
\ref{Figure::SymmetricMode}(A) has several distinctive features.
First, the hole at the location of the cavity is reduced in radius
to $r_c=0.21/a$. The nearest neighboring holes were also reduced
in radius, to approximately $r_h=0.26/a$, as well as outwardly
displaced from their original locations by $\sim 0.15/a$.
Qualitative arguments for these features, which were also observed
by Vuckovic, \textit{et al.}, have been
proposed\cite{Vuckovic2001}. Reducing the size of the defect hole
draws a bulk mode from the \textit{air band} into the photonic
band gap.  Air band modes are characterized by higher intensities
in regions where the index of refraction is lower, such as in the
air hole.  Therefore, the reduced cavity radius is most likely the
result of the design requirement that maximizes the intensity at
$\mathbf{r}_a=0$.

Decreasing the radii of the neighboring holes and moving them away
from the cavity reduces the intensity of the secondary lobes
surrounding the main peak in the cavity mode [see Figure
\ref{Figure::SymmetricMode}(B)]. These lobes, which would normally
coincide with air holes in the hexagonal lattice, experience an
atypically high index of refraction.  The lobe intensities are
suppressed because the defect mode, which was pulled from the air
band, is low-index seeking.  As a result, the cavity mode displays
better localization.  Therefore, the displaced neighboring holes
likely result from the mode volume minimization design
requirement.

\begin{figure}[ht]
\caption{Dispersion relationship and bulk mode contributions to
the optimized cavity mode for a radially symmetric defect.  The
upper shaded region denotes the free space light line.  The band
structure (solid lines) of the optimized photonic crystal is shown
and the points indicate bulk modes that contribute more than 1\%
to the cavity mode.  The dotted lines represent the band structure
of a photonic crystal with air holes.
\label{Figure::SymmetricDispersion}}
\end{figure}

By far, the most significant feature of the optimized photonic
crystal in Figure \ref{Figure::SymmetricMode} is that the
\textit{index of refraction of the bulk holes is increased} to
$n_h\sim 1.9$.  The optimal structure is a bulk photonic crystal
whose holes are made of a material other than air. Understanding
this result requires analyzing how the cavity mode is constructed
from the bulk mode basis functions.  The solid lines in Figure
\ref{Figure::SymmetricDispersion} show the dispersion relationship
of the optimized crystal (with non-air holes).  For reference, the
dispersion relationship for a corresponding crystal with air holes
($n_h=1.0$) is denoted by the broken lines. The points (circles)
represent bulk modes, identified by their band index, wavevector
and frequency, that contribute more than $1\%$,
$|a^{(m)}_{n,\mathbf{q}}|^2>10^{-2}$, to the optimized cavity
mode.

As can be seen in Figure \ref{Figure::SymmetricDispersion}, the
crystal with $n_h=1.9$ contains more bands that lie below the
light line.  Consequently, more modes, particularly with larger
wavevectors and frequencies, contribute to the cavity mode without
sacrificing Q (since these modes now lie below the light line).
Most likely, the ability to include contributions from more bulk
crystal modes allows an increase in the cavity Q while
simultaneously decreasing the mode volume.

The rationale here is that leaky modes must be excluded by the Q
maximization. However, constructing a cavity mode that is well
localized, \textit{i.e.}, not periodic, requires a large number of
Fourier components, including high frequency modes.  Drawing bands
out of the light cone provides access to more basis functions and
this allows a localized cavity field to be constructed without
resorting to leaky modes. Still, the optimization did incorporate
several bulk modes from above the light line, particularly from
the $J-\Gamma$ line.

We do not attempt to comment on how photonic crystals with non-air
holes might be fabricated.  However, it is important to consider
the major effects of adjusting the hole dielectric.  Increasing
the hole index of refraction decreases the band gap. However,
because the index contrast between the bulk material and the holes
is smaller, it should be possible to increase the hole radius
without suffering as much vertical scattering from the edges.  Of
course, increasing the hole size pushes bands back into the light
cone and a balance must be found.

\subsection{Illustration: Asymmetric Defect}
   \label{Section::AsymmetricDefect}

As a second demonstration, we relaxed the radial symmetry
requirement, and considered an arbitrary defect in a
two-dimensional hexagonal lattice. The same photonic crystal
parameters from Section \ref{Section::SymmetricDefect} were
adopted: a bulk index of refraction, $n_b=3.4$, and hole radius,
$r_h/a=0.3$.  Again, the mode expansion was constructed using five
photonic crystal layers surrounding the defect center to provide a
sufficient bulk mode basis expansion without incurring excessive
computational expense.

\begin{figure}[ht]
\caption{ The photonic crystal (A) and the cavity mode (B) that
result from optimizing the Q, mode volume, and cavity intensity by
solving a two-dimensional inverse problem (radial symmetry not
imposed). The cavity hole was reduced in radius to $r_c=2.18/a$,
and the neighboring holes were elongated in the vertical
direction. The most significant feature of the optimized structure
is that the index of refraction of the bulk crystal holes was
increased to $n_h=1.75$. \label{Figure::AsymmetricMode}}
\end{figure}

\begin{figure}[hb]
\caption{Dispersion relationship and bulk mode contributions to
the optimized cavity mode for a radially symmetric defect.  The
upper shaded region denotes the free space light line.  The band
structure (solid lines) of the optimized photonic crystal is shown
and the points indicate bulk modes that contribute more than 1\%
to the cavity mode.  The broken lines represent the band structure
of a photonic crystal with air holes.
\label{Figure::AsymmetricDispersion}}
\end{figure}

Solving the linear system of inversion equations was performed by
nesting the optimization within a conjugate-gradient search for
the best $\beta_i$ weighting parameters.  The photonic crystal was
again constructed by taking the contour level half way between the
minimum and maximum of $\eta(\mathbf{r})$ to eliminate the
numerical effects of truncating the Fourier expansions.

The resulting photonic crystal and cavity mode is depicted in
Figure \ref{Figure::AsymmetricMode}.  Here, the volume of the 2D
mode was reduced to $\sim$$\lambda^2/10$.  The center photonic
crystal hole was reduced in radius to $r_c=0.22/a$ and remained
circular. However, the nearest neighbor holes were deformed mainly
in the $\hat{y}$ direction. The minor axes, along $\hat{x}$, of
the four holes above and below the cavity were reduced to
$r_h~\sim0.26/a$. Their corresponding major axes were
simultaneously increased to $0.39/a$, resulting in the elliptical
holes seen in Figure \ref{Figure::AsymmetricMode}(A). These four
neighboring holes were also radially displaced from their normal
hexagonal lattice sites by $0.05/a$, smaller than what was
observed for the symmetric defect.

There was also a deformation of holes lying along the
$\hat{x}$-axis.  These were elongated in the $\hat{y}$ direction;
however, the degree of eccentricity was not constant. The minor
axes of the two horizontally neighboring holes decreased to
$0.26/a$, while their major axes increased to $0.31/a$.  The
remaining horizonal holes were also vertically stretched, with
major axes given by $.36/a$, $.31/a$ and $.30/a$ for photonic
crystal layers 3, 4 and 5, respectively.  Slight deformations
along the $\hat{x}$ direction were observed in the $\hat{y}$-axis
holes. However, their deformation was small, with major axes of
$0.31/a$ along the $\hat{x}$ direction, and minor axes of
$0.29/a$.  No consistent deformation of hole size was observed for
any other direction.

The qualitative arguments suggested by Vuckovic, \textit{et
al.}\cite{Vuckovic2001}, also apply to this structure. Replacing
air hole sites with higher index material suppresses the amplitude
of air-band originated modes. But more importantly, the partial
elongation of holes along the $\hat{x}$ axis bears significant
resemblance to the ``fractional edge delocations'' suggested by
the Scherer group\cite{Vuckovic2000, Vuckovic2001}.

As with the radially symmetric defect, the index of refraction of
the bulk holes increased, but only to $n_h\sim 1.8$.  A similar
argument based on pulling higher bands out of the light cone again
applies.  However, it is not as easy to provide an argument for
the elongated holes surrounding the defect and along the $\hat{x}$
and $\hat{y}$ axes.   Nonetheless, it can be seen from Figure
\ref{Figure::AsymmetricMode}(B) that better mode localization was
achieved by allowing an asymmetric cavity defect.

Figure \ref{Figure::SymmetricDispersion} shows the band diagram
for the optimized photonic crystal.  The dispersion relationship
for bulk holes with $n_h=1.8$ is depicted by the solid lines, and
the air hole bands are given by the broken lines.  The circles
represent bulk modes that contribute more than $1\%$,
$a^{(m)}_{n,\mathbf{q}}>10^{-2}$, to the optimized cavity mode.
The same argument used to explain the symmetric defect results can
be made here.  The crystal with $n_h=1.8$ contains more bands
below the light line.

A distinctive feature of the optimal asymmetric mode is that it
contains fewer contributions from leaky modes (inside the light
cone) than for the symmetric case. This can be seen by comparing
Figures \ref{Figure::SymmetricDispersion} and
\ref{Figure::AsymmetricDispersion}.   Exactly how the larger Q was
achieved for the asymmetric cavity is not clear. However, it is
not surprising that constraining the optimization (for example by
imposing a symmetry restriction) reduces the ability of the
inversion to satisfy the design requirements.

In these two-dimensional examples, solving the design problem by
inversion resulted in structures that differ from the previous,
trial and error designs that have been suggested.  These new
photonic crystal structures, however, might prove difficult to
fabricate. As such, they provide an indication of the performance
of an ideal, perhaps experimentally impractical, cavity.

\begin{figure}[t]
\caption{Schematic of the planar photonic crystal optimized using
numerical inversion methods.  $a$ is the hexagonal lattice
constant, $d$ is the thickness of the slab and $r_h$ is the radius
of the bulk crystal holes. The design parameters were obtained by
computing the crystal electromagnetic fields produced by plane
wave illumination.  The cavity Q was obtained by computing the
reflection coefficient, $R(\omega)$ as a function of incident wave
frequency. \label{Figure::PlanarLattice}}
\end{figure}

\section{Numerical Inversion Results} \label{Section::Numerical}

As a final example of inverting a photonic crystal defect, we
considered a hexagonal structure with finite depth to enable a
direct calculation of the cavity Q.  Additionally, we constrained
the index of refraction of the crystal holes to $n_h=1$.  In this
demonstration, the aim was to identify an optimal planar photonic
structure without requiring a more complicated fabrication.

For the planar structure, it was necessary to abandon an
analytical solution and perform the inversion optimization
numerically.  In some respects, the numerical design problem was
much simpler than the analytic case--- it was possible to maximize
the inversion cost functional, $\mathcal{J}$, without the need for
a specific mathematical analysis. Instead, the spatial dependence
of the dielectric function was parameterized.  Then the inverse
problem was solved by optimizing $\mathcal{J}[\eta(\mathbf{r})]$
over the parameter space using a genetic algorithm.

However, numerical methods introduce several new challenges. For
example, since the inversion cost functional involved a large
number of parameters, the resulting multi-dimensional optimization
was complicated by the existence of local minima. Additionally,
numerically integrating the Maxwell equations is computationally
expensive.  For instance, a single mode volume and cavity Q
calculation can require several minutes of computer time.

\begin{figure}[b]
\caption{Optimization progress as a function of genetic algorithm
generation number for the numerical inversion involving a planar
photonic crystal. \label{Figure::Optimization}}
\end{figure}

\subsection{Planar Photonic Crystal}
   \label{Section::PlanarInversion}

We considered an eight layer, two-dimensional hexagonal lattice
slab, as depicted in Figure \ref{Figure::PlanarLattice}.  Based on
the previous arguments (\textit{c.f.}, Section
\ref{Section::SymmetricDefect}), the hole radius was chosen to be
$r_h/a=0.3$ with a bulk material index of refraction, $n_b=3.4$.
The slab thickness was taken to be, $d/a=3/4$ in order to prevent
multimode behavior\cite{Loncar2000}.  Parameterizing the
dielectric function was accomplished by treating the positions and
major and minor axes of the lattice holes as optimization
variables. Specifically, the cavity hole, the holes surrounding
the defect, and the $\hat{x}$-axis holes were considered.  In all,
there were 52 optimization variables, 4 for 13 different lattice
sites.

\begin{figure}[b]
\caption{The
photonic crystal (A) and the cavity electric field
   (B) that result from optimizing the Q, mode volume, and peak
   cavity intensity by performing a numerical inversion for
   a planar photonic crystal.
\label{Figure::PlanarMode}}
\end{figure}

The Maxwell equations were solved by employing the transfer matrix
method (TMM) of Pendry and Bell\cite{Pendry1996}.  This technique
provided a means for computing the stationary states of the
electromagnetic field for a crystal illuminated on one edge by a
plane wave (refer to Figure \ref{Figure::PlanarLattice}).  The
spatial dependence of the electromagnetic fields, as well as the
crystal transmission and reflection coefficients were calculated
using a first order finite difference solution to Maxwell's
equations. The integration mesh used to compute the wavefields
extended five layers above and below the surface of the slab and
absorbing boundary conditions were imposed at the top and bottom
of the integration cell.

The mode volume was computed by integrating the square magnitude
of the electric field over the interior of the photonic crystal.
The volume integration was performed using a three dimensional
Simpson's rule quadrature, and the result was scaled by the field
maximum, according to Eq.\ (\ref{Equation::ModeVolume}). The
cavity Q was directly computed by scanning the reflection
coefficient over the frequency band gap.  The full width at half
maximum (FWHM) of the reflection lineshape was then used to
determine the cavity Q.

The inversion cost functional was optimized by employing a genetic
algorithm\cite{Goldberg} (GA) to maximize
$\mathcal{J}[\eta(\mathbf{r})]$.  A GA was chosen because,
although not terribly efficient, genetic algorithms provide good
exploration to exploitation in multi-dimensional searches.  As
such, they are generally successful at avoiding local minima on a
complex optimization surface.  However, good exploration does not
come without cost.  27 GA generations (iterations of the
algorithm) were required to optimize $\mathcal{J}$, as can be seen
Figure \ref{Figure::Optimization}.  With a population size of 10,
a steady-state propagation routine, a mutation rate of $15\%$ and
a crossover rate of $85\%$, the full optimization required 24.5
hours of CPU time on an Intel PIII 1.2 GHz desktop machine.

The resulting photonic crystal and the optimized cavity mode are
shown in Figure \ref{Figure::PlanarMode}.  As can be seen, a
similar structure to the two-dimensional optimization was
obtained.   The cavity defect hole was reduced in radius to
$0.21/a$ and the holes surrounding the defect were also decreased
in size.  The holes along the $\hat{x}$-axis were also elongated
in the $\hat{z}$ direction, which again demonstrated remarkable
similarity to the fractional edge delocations suggested by the
Scherer group\cite{Vuckovic2001,Vuckovic2001a}.

The mode volume of the field in Figure \ref{Figure::PlanarMode}
was found to be $\mathcal{V}_m \sim \lambda^3/3$ and the cavity Q
was $\sim$$1.1\times 10^5$.  These quantities surpass those of the
best PBG cavity designs to date.

\section{Conclusion} \label{Section::Conclusion}

We demonstrated that inversion methods provide a powerful
technique for designing photonic crystals for cavity QED
experiments. Both an analytical solution for a two-dimensional
crystal and numerical results for a planar (finite thickness)
crystal were presented.  In both cases, the design objectives,
namely high cavity Q and small mode volume, were achieved.

For two-dimensional crystals, it was shown that contributions to
the cavity mode from leaky bulk modes could be minimized without
delocalizing the field.   Here, inversion produced design
alternatives not previously suggested by trial and error, or
intuitive design.   The optimal cavity defect for both the
radially symmetric and asymmetric defect optimizations called for
a lattice of holes with an index of refraction greater than one.
An explanation for this inversion result, \textit{i.e.}, bulk
crystal holes composed of a material other than air, was
suggested. Increasing the index of refraction of the lattice holes
pulls more of the band diagram out of the light cone.
Consequently, less of the higher frequency bulk modes are leaky.
It was argued that a better localized cavity field can be
constructed by incorporating these higher frequency basis
functions.

The inverse problem design approach was also applied to a planar
crystal in order to treat a more realistic structure. However,
this required that numerical methods for integrating the Maxwell
equations as well as optimizing the inversion cost function be
adopted.  It was demonstrated that a crystal with low mode volume
and high Q could be achieved.  In order to minimize the effects of
local minima in the inversion optimization, a genetic algorithm
was adopted, despite its computational expense.

For the planar photonic crystal inversion, the dielectric constant
of the bulk lattice holes was constrained to $n_h=1$.  This
provided an example of restricting the inversion optimization
because of fabrication considerations.  Although it may be
possible to manufacture the inversion designs from Section
\ref{Section::Analytical}, \textit{i.e.}, bulk lattice holes with
$n_h>1$, it is also important to identify the best possible
structure which can fabricated using currently available
techniques.

An aspect of the design process that was not considered in this
paper was robustness.  Ideally, the photonic crystal would be
insensitive to slight variations in its structure, as might result
from small fabrication errors, temperature fluctuations, and
\textit{etc}.  Including robustness into design optimization
problems would add additional constraints into the optimization
process, however, algorithmic approaches to finding robust optima
are known\cite{Dullerud,Geremia4}. In addition to investigating
robustness, we plan to explore the use of optimization methods to
design PBG structures with convenient geometric features, such as
enlarged holes for atom trapping.

It was shown that the mathematical analysis tools utilized here
simultaneously optimize multiple, complimentary design
requirements.  Achieving similar outcomes via trial and error
methods is generally a formidable task.  Therefore, an algorithmic
approach to design problems with multiple objectives, such as with
photonic band gap materials, will likely only be possible via an
algorithmic approach.

\begin{acknowledgments}
   This work was supported by the DoD Multidisciplinary University Research
   Initiative (MURI) program administered by the Army Research Office
   under Grant DAAD19-00-1-0374.
\end{acknowledgments}

\appendix
\section{2D Inversion Equations} \label{Appendix::Details}

In this appendix we describe in greater detail how to extract the
defect dielectric function from the optimized expansion
coefficients (refer to Section \ref{Section::2DInversion}). The
general procedure involves substituting the normal mode expansion,
Eq.\ (\ref{Equation::BulkModeBloch}), and the Fourier expansions,
Eqs.\ (\ref{Equation::BulkFourier}) and
(\ref{Equation::DefectFourier}), into the Maxwell equation, Eq.\
(\ref{Equation::MaxwellCurl}). The resulting inversion equation,
Eq.\ (\ref{Equation::2DSystem}), results from employing the
orthogonality of the bulk modes.

In doing the algebra, it is more convenient to work a bulk mode
expansion that includes wave vectors from all of k-space,
\textit{i.e.}, by incorporating multiple Brillouin zones,
\begin{equation}
   \mathbf{H}_m(\mathbf{r}) = \frac{1}{N}
   \sum_{n,\mathbf{k}} a^{(m)}_{n,\mathbf{k}}
   H_{n,\mathbf{k}}(\mathbf{r})
\end{equation}
where $N$ is the number of Brillouin zones (reciprocal lattice
vectors). The orthogonality relationships are now give by,
\begin{equation}
    \int_{V_N} \mathbf{H}^*_{n^\prime,\mathbf{k}^\prime}(\mathbf{r})
    \mathbf{H}_{n,\mathbf{k}}(\mathbf{r}) d\mathbf{r}
    = \delta_{n,n^\prime} \sum_{\mathbf{G}}
    \delta_{\mathbf{k}^\prime,\mathbf{k+G}}
\end{equation}
where the integration is over the $N$ associated Wigner-Seitz
cells.

Substituting both the bulk crystal and defect dielectric Fourier
expansions into the Maxwell curl equation,
\begin{eqnarray}
   & \nabla \times \eta_0(\mathbf{r}) \nabla \times
   \mathbf{H}_m(\mathbf{r}) & \nonumber \\
   & + \nabla \times \delta\eta(\mathbf{r}) \nabla \times
   \mathbf{H}_m(\mathbf{r})
   = \frac{\omega_m^2}{c^2} \mathbf{H}_m(\mathbf{r}) & \nonumber
\end{eqnarray}
left multiplying by, $\mathbf{H}_{n'',\mathbf{q}''}$, and
integrating leads to,
\begin{eqnarray}
 & \frac{1}{N} \sum\limits_{n,\mathbf{k}} a^{(m)}_{n,\mathbf{k}}
 \frac{\omega_{n,\mathbf{k}}^2 - \omega_m^2}{c^2}
 \Large\int_V \mathbf{H}^*_{n^{\prime\prime},\mathbf{k}^{\prime\prime}}(\mathbf{r})
 \mathbf{H}_{n,\mathbf{k}}(\mathbf{r}) d\mathbf{r}
 = & \nonumber \\
 & \frac{1}{N}\sum\limits_{\mathbf{k}}\sum\limits_{n,\mathbf{k}}
 \sum\limits_{\mathbf{G},\mathbf{G}^{\prime\prime}}
 a^{(m)}_{n,\mathbf{k}} \delta\eta_{\mathbf{k}}
 h^*_{n^{\prime\prime},\mathbf{k^{\prime\prime}+G^{\prime\prime}}} h_{n,\mathbf{k+G}} &\\
  & (\mathbf{k^{\prime\prime}+G^{\prime\prime}})
\cdot (\mathbf{k^{\prime\prime}+G^{\prime\prime}-k^\prime}) &
\nonumber \\
 & \int_V e^{i(\mathbf{k+G+k^\prime-k^{\prime\prime}-G^{\prime\prime}
 )\cdot\mathbf{r}}} d\mathbf{r}& \nonumber
\end{eqnarray}
where, again, the integrations are over the $N$ unit cells.
Applying the orthogonality of the bulk modes leads to a set of
equations that still run over all wavevectors,
\begin{eqnarray}
   &\frac{1}{N} \sum\limits_{n,\mathbf{k}}
      \sum\limits_{\mathbf{k}^\prime} \sum\limits_{\mathbf{G}''}
      a^{(m)}_{n,\mathbf{k}} \delta\eta_{\mathbf{k}'}
      h^*_{n'',\mathbf{k''+G''}} h_{n,\mathbf{G}''-\mathbf{k}'}
      & \\
   & (\mathbf{k''+G''})\cdot(\mathbf{k''+G''-k'})
    = \frac{-1}{N} a^{(m)}_{n'',\mathbf{k''}}
      \frac{\omega_{n'',\mathbf{k''}}^2 - \omega_m^2}{c^2} & \nonumber
\end{eqnarray}
Then, the summations can be folded back into the first Brillouin
zone by using the identity,
\begin{equation}
      \sum\limits_{n,\mathbf{k}} a^{(m)}_{n,\mathbf{k}}
      = N \sum\limits_{n,\mathbf{q}} a^{(m)}_{n,\mathbf{q}}
\end{equation}
and the indices can be renamed to produce Eq.\
(\ref{Equation::2DSystem}).

\end{document}